\documentclass[aps,prd,twocolumn,superscriptaddress,nofootinbib]{revtex4}
\usepackage{xspace} 
\usepackage{color}
\usepackage{graphicx}
\usepackage{amsmath,amssymb,latexsym}
\usepackage{bm}
\usepackage{epsfig}
\newcommand{\postscript}[2]{\setlength{\epsfxsize}{#2\hsize}
   \centerline{\epsfbox{#1}}}

\def\Offline{\mbox{$\overline{\textrm%
{Off}}$\hspace{.05em}\protect\raisebox{.4ex}%
{$\protect\underline{\textrm{line}}$}}\xspace}                                                                    

\newcommand{\md}{M_D}
\newcommand{\mbh}{M_{\text{BH}}}

\newcommand{\xmin}{x_{\text{min}}}

\newcommand{\cm}{\text{cm}}
\newcommand{\km}{\text{km}}
\newcommand{\g}{\text{g}}

\def\sh{\sqrt{\hat s}}
\def\be{\begin{equation}}
\def\ee{\end{equation}}
\begin{document}
\title{Using cosmic neutrinos to search for non-perturbative physics\\ at the Pierre Auger Observatory}
\author{Luis A.~Anchordoqui}
\affiliation{Department of Physics, 
University of Wisconsin-Milwaukee,
P.O. Box 413, Milwaukee, WI 53201, USA
}
\author{Haim Goldberg}
\affiliation{Department of Physics,
Northeastern University, Boston, MA 02115, USA
}
\author{Dariusz G\'ora} 
\affiliation{Karlsruhe Institute of Technology (KIT), 
D-76021 Karlsruhe, Germany
}
\affiliation{Institute of Nuclear Physics PAN, Krakow, Poland
}
\author{Thomas \nolinebreak Paul}
\affiliation{Department of Physics,
Northeastern University, Boston, MA 02115, USA
}
\affiliation{Karlsruhe Institute of Technology (KIT), 
D-76021 Karlsruhe, Germany
}
\affiliation{
University of Nova Gorica,
Vipavska 13, POB 301,
SI-5001 Nova Gorica, Slovenia 
}
\author{Markus Roth}
\affiliation{Karlsruhe Institute of Technology (KIT), 
D-76021 Karlsruhe, Germany
}
\author{Subir Sarkar}
\affiliation{Rudolf Peierls Centre for Theoretical Physics,
University of Oxford,
Oxford OX1 3NP, UK
}
\author{Lisa Lee Winders}
\affiliation{Department of Physics, 
University of Wisconsin-Milwaukee,
P.O. Box 413, Milwaukee, WI 53201, USA
}
\begin{abstract}
\noindent 
The Pierre Auger (cosmic ray) Observatory provides a laboratory for
studying fundamental physics at energies far beyond those available at
colliders. The Observatory is sensitive not only to hadrons and
photons, but can in principle detect ultrahigh energy neutrinos in the
cosmic radiation. Interestingly, it may be possible to uncover new
physics by analyzing characteristics of the neutrino flux at the
Earth. By comparing the rate for quasi-horizontal, deeply penetrating
air showers triggered by all types of neutrinos, with the rate for
slightly upgoing showers generated by Earth-skimming tau neutrinos, we
determine the ratio of events which would need to be detected in order
to signal the existence of new non-perturbative interactions beyond
the TeV-scale in which the final state energy is dominated by the
hadronic component.  We use detailed Monte Carlo simulations to
calculate the effects of interactions in the Earth and in the
atmosphere. We find that observation of 1 Earth-skimming and 10
quasi-horizontal events would exclude the standard model at the 99\%
confidence level. If new non-perturbative physics exists, a decade or
so would be required to find it in the most optimistic case of a
neutrino flux at the Waxman-Bahcall level and a neutrino-nucleon
cross-section an order of magnitude above the standard model
prediction.
\end{abstract}
\maketitle
\section{Introduction}

Ultrahigh energy cosmic neutrinos (UHEC$\nu$) are expected to be
produced in association with the observed ultrahigh energy (charged)
cosmic rays (UHECR), either at the same sites responsible for UHECR
acceleration, or via interaction of the UHECR during propagation,
particularly with the cosmic microwave background (CMB). These
neutrinos are unique probes of new physics as their interactions are
uncluttered by the strong and electromagnetic forces and, upon arrival
at the Earth, they may experience collisions with center-of-mass
energies up to $\sqrt{s} \alt 250$~TeV.  However, rates for new
physics processes are difficult to test since the flux of cosmic
neutrinos is virtually unknown.  Interestingly, it is possible in
principle to disentangle the unknown flux and new physics processes by
using multiple
observables~\cite{Kusenko:2001gj,Anchordoqui:2001cg,Anchordoqui:2005ey,PalomaresRuiz:2005xw,Anchordoqui:2006ta}.

The Pierre Auger Observatory provides a promising way to detect
UHEC$\nu$ by looking for deeply--developing, large zenith angle ($\agt
75^\circ$) or ``quasi-horizontal'' air
showers~\cite{Capelle:1998zz}. At these large angles, hadron-induced
showers traverse the equivalent of several times the depth of the
vertical atmosphere and consequently their electromagnetic component
is extinguished before reaching the detector.  Only very high energy
muons survive past about 2 equivalent vertical atmospheres. Therefore,
the shape of a hadron-induced shower front is very flat and prompt in
time. In contrast, a neutrino shower exhibits the roughly same
morphology as a vertical shower.  It is therefore possible to
distinguish neutrino induced events from background hadronic
showers. Moreover, because of full flavor mixing, tau neutrinos are
expected to be as abundant as other species in the cosmic
flux. $\nu_\tau$'s can interact in the Earth's crust, producing $\tau$
leptons which may decay above to the Auger detectors; such events will
be referred to as ``Earth--skimming''
events~\cite{Feng:2001ue,Bertou:2001vm}.

Possible deviations of the neutrino--nucleon cross-section due to new
non-perturbative interactions\footnote{Throughout this paper we use
  this term to describe neutrino interactions in which the final state
  energy is dominated by the hadronic component. We are {\em not}
  considering here new ``perturbative'' physics e.g. (softly broken)
  supersymmetry at the TeV scale which would have quite different
  signatures in cosmic ray showers.} can be uncovered at the Auger
Observatory by combining information from Earth-skimming and
quasi-horizontal showers.  In particular, if an anomalously large rate
is found for deeply developing quasi-horizontal showers, it may be
ascribed either to an enhancement of the incoming neutrino flux, or an
enhancement in the neutrino-nucleon cross-section (assuming
non-neutrino final states dominate).  However, these possibilities can
be distinguished by comparing the rates of Earth-skimming and
quasi-horizontal events.  For instance, an enhanced flux will increase
both quasi-horizontal and Earth-skimming event rates, whereas an
enhanced interaction cross-section will also increase the former but
{\em suppress} the latter, because the hadronic decay products cannot
escape the Earth's crust.  Essentially this approach constitutes a
straightforward counting experiment, as the detailed shower properties
are not employed to search for the hypothesized new physics.

In this paper, we compute how many Earth-skimming {\it vs.}
quasi-horizontal showers one would have to collect at the Auger
Observatory to convincingly demonstrate new non-perturbative physics
in which the final state energy is dominated by the hadronic
component.  We show that even a small number of events could be
sufficient to rule out the standard model (SM).  Thus the expected low
neutrino ``luminosity'' is not at all a show-stopper, and the
Observatory has the potential to uncover new physics at scales
exceeding those accessible to the LHC.

In order to demonstrate this, we first compute acceptances for
Earth-skimming and quasi-horizontal events using detailed models of
the terrain in the vicinity of the Observatory as well as detailed
simulations of the response of the Surface Array to highly inclined
air showers.  We then perform a likelihood analysis to determine the
event counts needed to exclude the SM at various confidence
levels. The analysis includes systematic effects both from theoretical
uncertainties in the (perturbative QCD) SM
cross-section~\cite{CooperSarkar:2007cv} and from uncertainties
associated with the detector response.

The outline of the paper is as follows.  In Sec.~\ref{sec:new} we
discuss some possible new physics scenarios which could manifest
themselves as non-perturbative interactions at LHC energies and
beyond.  Next, in Sec.~\ref{sec:sims} we describe the detailed Monte
Carlo studies of the acceptance for Earth-skimming and
quasi-horizontal showers under the assumption of SM interactions,
including systematic uncertainties.  Finally in Sec.~\ref{sec:results}
we perform the statistical analysis to ascertain the discovery reach
of the Observatory. Our conclusions are collected in
Sec.~\ref{summary}

\section{New Non-Perturbative Physics at the LHC Energy Scale and
  Beyond} \label{sec:new}

The analysis techniques described herein constitute an entirely
general approach to searching for non-perturbative interactions in
which the final state is dominated by hadrons, without any dependence
on what hypothetical mechanism might actually cause the
`hadrophilia'. In order to illustrate some possible new physics
signals which may be accessible using these techniques, we consider
below two interesting possibilities.

\subsection{TeV-scale mass black holes}

In $D$-dimensional scenarios with large-compact-extra-dimensions (of common linear size $2\pi r_{\rm c}$) the Planck scale is related to the fundamental scale of gravity ($M_D$) according to~\cite{Arkani-Hamed:1998rs} \begin{equation}
 M_{\rm Pl}^2 = 8 \pi \,r_{\rm c}^{D-4} M_D^{D-2} \,\, .
\end{equation} 
If $M_D \agt M_W = G_{\rm F}^{-1/2} \simeq 300$~GeV, microscopic black
holes (BH) can be produced {\em gravitationally} in particle
collisions with center-of-mass energies $s \agt 1~{\rm
  TeV}$~\cite{Banks:1999gd}. Subsequently a TeV-scale BH would
promptly decay via thermal Hawking radiation~\cite{Hawking:1974rv}
into observable quanta~\cite{Emparan:2000rs}. (For $M_D = 1$~TeV, the
lifetime of a BH of mass 10 TeV is less than $10^{-25}$~s.) Since
gravitational coupling is flavor blind, a BH emits all the $\approx
120$ SM particle and anti-particle degrees of freedom with roughly
equal probability. Accounting for color and spin, we expect $\approx
75\%$ of the particles produced in BH evaporation to be quarks and
gluons, $\approx 10\%$ charged leptons, $\approx 5\%$ photons or $W/Z$
bosons, and $\approx 5\%$ neutrinos.  Thus, TeV BH production and
evaporation constitutes a clear example of beyond-SM non-perturbative
physics.

Although such BH production cross-section $\sim M_W^{-1}$ is 5 orders
of magnitude smaller than the QCD cross-section $\sim \Lambda_{\rm
  QCD}^{-1}$, it was proposed \cite{Dimopoulos:2001hw} that such BHs
could be produced copiously at the LHC, and that these spectacular
events could be easily filtered out of the QCD background.  This is
possible by triggering on BH events with prompt charged leptons and
photons, each carrying hundreds of GeV of energy.

Cosmic ray collisions, with center-of-mass energies ranging up to
$10^5$~GeV, certainly produce BHs if the LHC does. The question is,
can they be detected?  Most cosmic rays are protons or heavier nuclei,
which collide with hadrons in the upper atmosphere, producing
cascading showers which eventually reach the Earth's surface.  At
energies of interest, however, the cosmic ray luminosity ($L \sim 7
\times 10^{-10}~(E/{\rm PeV})^{-2}$~cm$^{-2}$ s$^{-1}$, taking a
single nucleon in the atmosphere as a target and integrating over $2
\pi$ sr), is about 50 orders of magnitude smaller than the LHC
luminosity, thus making it futile to hunt for BHs in hadronic cosmic
ray interactions.  On the other hand, neutrino interaction lengths are
far longer than the Earth's atmospheric depth, although they would be
greatly reduced by the cross-section for BH
production~\cite{Feng:2001ib}. Cosmic neutrinos therefore could
produce BHs with roughly equal probability at any point in the
atmosphere.  As a result, the light descendants of the BH may initiate
low-altitude, quasi-horizontal showers at rates significantly higher
than SM predictions.

Analytic and numerical studies have revealed that gravitational
collapse takes place at high energies and small impact
parameters~\cite{Eardley:2002re,Yoshino:2002tx}. In the course of
collapse, a certain amount of energy is radiated in gravitational
waves, leaving a fraction $y \equiv \mbh/\sh$ available for Hawking
evaporation. Here, $\mbh$ is a {\it lower bound} on the final mass of
the BH and $\sh = 2 x m_N E_\nu$ is the center-of-mass energy of the
colliding particles, taken to be partons. This ratio depends on the
impact parameter of the collision, as well as on the dimensionality of
space-time.

The inclusive production of BHs proceeds through different final
states for different classical impact parameters
$b$~\cite{Yoshino:2002tx}. These final states are characterized by the
fraction $y(z)$ of the initial $\sqrt{\hat s}$ which is trapped within
the horizon. Here, $z= b/b_{\rm max},$ and $b_{\rm max}= \sqrt{F} \,
r_s$ is the maximum impact parameter for collapse, where
\begin{equation}
\label{schwarz}
r_\mathrm{s} =
\frac{1}{\md}
\left[ \frac{\sqrt{\hat s}}{\md} 
\, \frac{2^{D-4} \pi^{(D-7)/2}\Gamma({D-1\over 2})}{D-2}
\right]^{\frac{1}{D-3}}
\end{equation}
is the radius of a $D$-dimensional Schwarzschild BH~\cite{Myers:un},
and $F$ is the form factor~\cite{Yoshino:2002tx}.

The $y$ dependance complicates the parton model calculation, since the
production of a BH of mass $M_{\rm BH}$ requires that $\hat s$ be
$M_{\rm BH}^2/y^2(z)$, thus requiring the lower cutoff on parton
momentum fraction to be a function of impact parameter.  Because of
the complexity of the final state, we assume that amplitude
intereference effects can be ignored and we take the $\nu N$
cross-section as an impact parameter-weighted average over partonic
cross-sections, with the lower parton fractional momentum cutoff
determined by $x_{\rm min} = M_{\rm BH}^{\rm min}/M_D$.  This gives a
lower bound ${\cal X} = (\xmin M_D)^2/[y^2(z)s]$ on the parton
momentum fraction $x$. All in all, the $\nu N \to {\rm BH}$
cross-section reads~\cite{Anchordoqui:2003jr}
\begin{equation}
\sigma (\nu N \to {\rm BH}) =  \int_0^1 2 z \,dz 
\int_{{\cal X}}^1 dx \, F \pi r_s^2 \,\sum_i f_i(x,Q) \ ,
\label{sigma}
\end{equation}  
where $i$ labels parton species and the $f_i(x,Q)$ are parton
distribution functions (PDF).

As an illustration, we consider the $D=10$ string inspired
scenario. For, $M_D = 1$~TeV, $x_{\rm min} = 1$, and primary neutrino
energy $E_\nu = 10^{10}$~GeV, we obtain $\sigma (\nu N \to {\rm
  BH})\sim 2 \times 10^6~{\rm pb}$~\cite{Anchordoqui:2003jr}. This is
about two orders of magnitude above SM predictions. The BH production
cross-section by UHEC$\nu$ scales as
\begin{equation}
\sigma(\nu N \to {\rm BH}) \propto 
 \left[\frac{1}{M_D^2}\right]^{\frac{D-2}{D-3}}  \,  .
\end{equation}
A further suppression arises if $x_{\rm min}$ is increased.  For
parameters in the semiclassical regime ($x_{\rm min} \agt
3$~\cite{Anchordoqui:2003ug}) the BH cross-section becomes comparable
to the SM cross-section at $M_D \sim 2$~TeV; this determines the
multidimensional Planck scale to which Auger may be
sensitive. However, the LHC will also be sensitive to
extra-dimensional effects at a similar scale.  It is interesting to
consider whether Auger may have access to new physics {\em beyond} the
reach of the LHC and we now discuss such a possibility.

\subsection{Sphalerons}

In the electroweak theory, non-trivial fluctuations in $SU(2)$ gauge
fields generate an energy barrier interpolating between topologically
distinct vacua~\cite{'tHooft:1976up}.  An index theorem describing the
fermion level crossings in the presence of these fluctuations reveals
that neither baryon nor lepton number is conserved during the
transition, but only the combination $B-L.$ Inclusion of the Higgs
field in the calculation modifies the original instanton
configuration~\cite{Klinkhamer:1984di}.  An important aspect of this
modification (called the ``sphaleron'') is that it provides an
explicit energy scale of about 10 TeV for the height of the barrier.
This barrier can be overcome through thermal transitions at high
temperatures~\cite{Kuzmin:1985mm}, providing an important input to any
calculation of cosmological baryogenesis. More speculatively, it has
been suggested~\cite{Aoyama:1986ej} that the topological transition
could take place in two particle collisions at very high energy. The
anomalous electroweak contribution to the partonic process can be
written as \begin{equation} \hat\sigma_i(\hat s)= 5.3\times 10^3 {\rm
    mb} \cdot e^{-(4\pi/\alpha_W)\ F_W(\epsilon)}\ \
  , \label{sigmahat} \end{equation} where $\alpha_W\simeq 1/30$, the
tunneling suppression exponent $F_W(\epsilon)$ is sometimes called the
``holy-grail function'', and $\epsilon \equiv \sqrt{\hat s} /(4\pi
m_W/\alpha_W)\simeq \sqrt{\hat s}/30$~TeV. Thus, it is even possible
that at or above the sphaleron energy the cross-section could be of
${\cal O}(\rm mb)$ \cite{Ringwald:2003ns}.
Of particular interest to cosmic ray physicists would be enhancement
of the neutrino cross-section over the perturbative SM estimates, say
by an order of magnitude in the energy range $9.5 < \log_{10}
(E_\nu/{\rm GeV}) < 10.5$. With the methods outlined in this paper,
this can be detected as an anomalous ratio of quasi-horizontal
Earth-skimming showers to upcoming showers.

It was shown~\cite{Ringwald:2003ns} that for the simple sphaleron
configuration $s$-wave unitarity is violated for $\sqrt{\hat s}> 4\pi
M_W /\alpha_W\sim 36$~TeV. For lower parton subenergies, the
cross-section is exponentially damped to values well below the
perturbative SM electroweak
value~\cite{Ringwald:2003ns,Bezrukov:2003er}.  On the other hand, at
the higher parton subenergies, the cross-section may well be dominated
by non-spherically symmetric classical field
configurations~\cite{Gould:1993hb}.  If for $\sqrt{\hat s} > 36$ TeV
we saturate unitarity in each partial wave then this yields a
geometric parton cross-section $\pi R^2$, where $R$ is some average
size of the classical configuration. As a fiducial value we take the
core size of the Manton-Klinkhamer sphaleron, $R\simeq 4/M_W\simeq
10^{-15} $~cm. In this simplistic model, the $\nu N$ cross-section is
\begin{equation}
\sigma^{\nu N}_{\rm black \, disk} (E_\nu) 
 = \pi R^2\ \int_{x_{\rm min}}^1 \sum_{partons} f(x)\ dx \,,
\end{equation}
where $x_{\rm min} = \hat s_{\rm min}/ s = (36)^2/ 2ME_\nu \simeq
0.065$.  In the region $0.065 < x < 3(0.065)$ the PDF for the up and
down quarks is well approximated by $f\simeq 0.5/x$, so the expression
for the cross-section becomes
\begin{eqnarray}
\sigma^{\nu N}_{\rm black \, disk} (E_\nu) 
&\simeq &\pi R^2\,\, (0.5)\,\, (\ln 3)\,\,  (2/2) \nonumber \\
& \simeq & 1.5\times 10^{-30} {\rm cm}^2  \,,
\end{eqnarray}
where the last factor of 2/2 takes into account the (mostly) 2
contributing quarks $(u,d)$ in this range of $x,$ and the condition
that only the left-handed ones contribute to the scattering. This is
about 80 times the SM
cross-section~\cite{CooperSarkar:2007cv,Gandhi:1998ri}. Of course this
calculation is very approximate and thecross-section can easily be
smaller by a factor of 10 (e.g., if $R$ is 1/3 of the fiducial value
used).

\section{Acceptance and Systematic Uncertainties}
\label{sec:sims}

To calculate the acceptance we perform detailed Monte Carlo
simulations.  The incoming neutrinos are propagated through the
Earth's crust, Andes mountains, and the atmosphere using an extended
version~\cite{Gora:2007nh} of the code ANIS
~\cite{Gazizov:2004va}. For fixed neutrino energies, $10^6$ events are
generated with zenith angles in the range $60^\circ-90^\circ$
(down-going showers) and $90^\circ-95^\circ$ (upgoing showers) and
with azimuth angles in the range $0^\circ-360^\circ$. Neutinos are
propagated along their trajectories of length $\Delta L$ from the
generation point on the top of the atmosphere to the detector in steps
of $\Delta L/1000 \, (\geq 6~{\rm km}).$ At each step of propagation,
the $\nu N$ interaction probability
\begin{equation}
  \label{prob}
  P (E_{\nu},E_{l},\theta)  \simeq N_{A} \ \sigma^{\nu N}_{\rm SM}(E_{\nu}) \
  \rho(Z)  \ \Delta L \ , 
\end{equation}
is calculated using the cross-section ($\sigma_{\rm SM}^{\nu N}
(E_{\nu})$) estimates of Ref.~\cite{CooperSarkar:2007cv}, where
$\rho(Z)$ is the local medium density, $E_l$ the energy of the
outgoing lepton, and $N_A \simeq 6.022 \times 10^{23}~\g^{-1}$. The
outcoming particle spectrum from $\nu N$ interactions is simulated
with PYTHIA~\cite{Sjostrand:2006za} and tau decays are simulated using
the package TAUOLA ~\cite{Jadach:1993hs}.

\begin{figure}[tpb]
\postscript{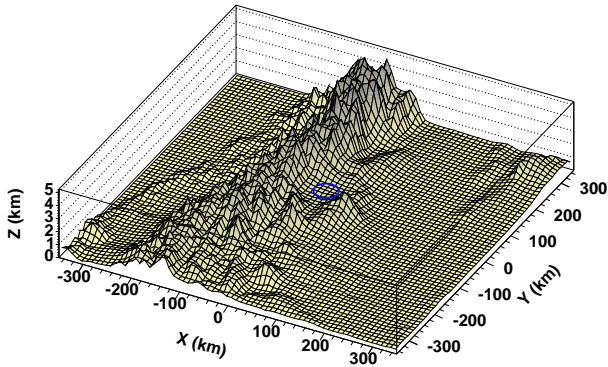}{0.99}
\caption{Topography of the Auger site according to CGIAR-CSI data.
  The center of the map corresponds to the centre of the Auger array
  (latitude $=35.25^{\circ}$~S, longitude $=69.25^{\circ}$~W). The
  Auger position is marked by a circle.}
\label{augermap}
\end{figure}

The flux of otutgoing leptons as well as their energy and the decay
vertex positions are calculated inside a defined detector volume. The
geometrical size of the detector volume is set to $3000 \times 10~{\rm
  km}^3$ and it includes the real shape of the Auger Observatory on
the ground. A relief map of the Andes mountains was constructed
according to a digital elevation data of the Consortium for Spatial
Information (CGIAR-CSI)~\cite{CGIAR-CSI}. The map of the area around
the Auger site is shown in Fig.~\ref{augermap}.

The detection volume corresponds to the so called {\em active volume}
in which potentially detectable neutrino interactions are
simulated. For a given incoming neutrino with energy $E_{\nu}$ the
active volume is defined by a particular plane $A_{\rm gen}$ and
distance $\Delta L$.  The plane $A_{\rm gen}$ is the cross-sectional
area of the detector volume and it is used as a reference plane for
the generation of incoming neutrinos.  The area depends on the zenith
angle $\theta$ of the incoming neutrino.  The distance $\Delta L$ is
the multiple, $n$, of the average lepton range $\langle
L^l(E_l)\rangle$.

Earth-skimming events occur in the Earth's crust, and so the relevant
neutrinos and taus sample only the Earth's surface density, $\rho_s
\approx 2.65~\g/\cm^3$.  At these energies, the tau's propagation
length is determined not by its decay length but by its energy loss.
The energy loss per unit length of crossed matter is usually
approximated by a linear equation (continuous energy loss
approach). The $\tau$ lepton loses energy in the Earth according to
\begin{equation}
\frac{dE_{\tau}}{dz} = -(\alpha_{\tau} + \beta_{\tau} E_{\tau})
\rho_s \ ,
\end{equation}
where the factor $\alpha_\tau$ parametrizes the ionisation losses and
$\beta_\tau$ the energy losses through bremsstrahlung, pair
production, and hadronic interactions. For $E_\nu = 10^7~{\rm GeV},$
$\alpha_\tau$ is negligible and $\beta_{\tau} \approx 0.8\times
10^{-6}~\cm^2/\g$~\cite{Dutta:2000hh}. Hadronic interactions 
(i.e., lepton-nucleus inelastic interactions dominated by small values
of the squared momentum transfer $Q^2$) are responsible for the
largest and the most uncertain
contribution~\cite{Armesto:2007tg}. Such an uncertainty in
$\beta_\tau$ dominates the systematic errors in the estimate of the
neutrino event rates.

To investigate the response of the Auger detector, we generate the
lateral profiles of the shower development using the output of PYTHIA
and/or TAUOLA as input for AIRES~\cite{Sciutto:2001dn}. The showers
induced by the products of up-going decaying tau leptons, with
energies from 0.1~EeV to 100~EeV and decay position at altitudes
ranging from 0 to 3500~m above sea leavel, are simulated in steps of
100~m. At each altitude 40 events are generated to cover the tau decay
channels implemented in ANIS~\cite{Gora:2007nh}. In the case of
down-going showers, the decay altitudes range from ground level up to
the upper atmosphere.

The response of the surface detector array is simulated in detail
using the \Offline simulation package~\cite{Argiro:2007qg}. Besides
the standard procedure to simulate the spacial and temporal signal
response we have added the simulation of atmospheric background muons
to study the impact on the neutrino identification, since such
accidental muons might be wrongly classified as shower particles.  The
background from hadronic showers above $10^8$~GeV is estimated to be
${\cal O} (1)$ in 20 years.  At $E_\nu \sim 10^{10}$~GeV the cosmic
ray flux is $\approx 10^6$ time smaller, so the expected background
for the energy bin considered here ($9.5 < \log_{10} (E_\nu/{\rm GeV})
< 10.5$) is negligible.

The expected neutrino event rate (of flavor $\alpha$) in the detector
volume is found to be
\begin{equation}
\label{nrate}
   N_{\nu_\alpha}=F^{w}_{\nu}   \, \sum_{i=1}^{N_{\rm acc}}  P_{i},
\end{equation}
where $N_{\rm acc}$ is the number of events triggering the detector and passing
all quality cuts of the cascade analysis. Here,
\begin{equation}
  \label{flux}
  F^{w}_{\nu}=  N_{\rm gen}^{-1}  \, \Delta T \,\int_{E_{\rm min}}^{E_{\rm max}}
  \Phi_0^{\nu_\alpha}(E_{\nu})  \, dE_{\nu}   \int_{\theta_{\rm min}}^{\theta_{\rm max}}
  A_{\rm gen}(\theta) \, d\Omega \,, 
\end{equation}
$d\Omega$ is the solid angle, $\Delta T$ the observation time, $N_{\rm
  gen}$ is the number of generated events from surface $A_{\rm gen}$,
and we take the neutrino flux $\Phi_0^{\nu_\alpha}(E_{\nu})$ to be
isotropic. We further assume $\nu_e : \nu_\mu : \nu_\tau \simeq
1:1:1$, which is generally thought to be the case if the neutrinos are
produced predominantly thorugh pion decay. In order to ascertain the
systematic uncertainties associated with our lack of knowlege of the
dependence of the flux on energy, we consider three scenarios which
plausibly bracket the range of possibilities:
\begin{enumerate}
\item $  \Phi_0^{\nu_\alpha}(E_\nu) = ( {\cal C}/E_0) \, E_\nu^{-1}$,
\item $ \Phi_0^{\nu_\alpha}(E_\nu) = {\cal C} \, E_\nu^{-2}$,
\item $\Phi_0^{\nu_\alpha}(E_\nu) = ( {\cal C}/E_0) \, E_\nu^{-3}$,
\item  $\Phi_0^{\nu_\alpha}(E_\nu) = {\cal C} E_\nu^{-2} \, 
{\rm exp} [-\log_{10} (E_\nu/E_0)^2/(2 \sigma^2)]$,
\end{enumerate}
where ${\cal C} = 2.33 \times 10^{-8}~{\rm GeV}\, {\rm s}^{-1}\, {\rm
  cm}^{-2}\, {\rm sr}^{-1}$, $E_0 = 10^{10}$~GeV, $\sigma =
0.5$~GeV. This normalization (2) constitutes the common benchmark, the
so-called `Waxman-Bahcall bound'~\cite{Waxman:1998yy}. The factor
$F^{w}_{\nu}$ of Eq.~\ref{flux} is chosen to yield the total number of
events per year.  The expected rates for the entire range over which
Auger is sensitive are given in Table~\ref{table1} and the rates for
the high energy bin considered in the following study are given in
Table~\ref{table2}.

\begin{table}[htb]
\vspace*{-0.1in}
\caption{Expected events per year ($N_i$) at Auger in the energy range $8 < \log_{10} (E_\nu/{\rm GeV})$, for various incident zenith angle ($\theta$) ranges, assuming the Waxman-Bahcall flux.}
\label{table1}
\begin{tabular}{c|@{}cc|@{}ccccc|c}
\hline
flux & \multicolumn{2}{@{}c|}{up-going}  & \multicolumn{5}{@{}c|}{down-going} & ratio \\
\hline
  & $\theta$  & $N_{\nu_\tau}$ & $\theta$  & $N_{\nu_e}$ & $N_{\nu_\tau}$  & $N_{\nu_\mu}$ & $N_{\nu_{\rm all} } $ & $N_\tau/N_{\nu_{\rm all}}$ \\
\cline{2-3} \cline{4-9}
$(2)$ &~90-95 & 0.68~&~60-90   & 0.134   & 0.109   & 0.019  & 0.262~& 2.58 \\
$(2)$ &~90-95 & 0.68~&~75-90 & 0.075 & 0.071 & 0.011 & 0.157~& 4.27 \\
\hline
\end{tabular}
\end{table}

\begin{table}[htb]
\vspace*{-0.1in}
\caption{Expected events per year ($N_i$) at Auger in the energy range $9.5 < \log_{10} (E_\nu/{\rm GeV}) < 10.5$, for various incident zenith angle ($\theta$) ranges and the 4 flux models considered.}
\label{table2}
\begin{tabular}{c|@{}cc|@{}ccccc|c}
\hline
flux & \multicolumn{2}{@{}c|}{up-going}  & \multicolumn{5}{@{}c|}{down-going} & ratio \\
\hline
  & $\theta$  & $N_{\nu_\tau}$ & $\theta$  & $N_{\nu_e}$ & $N_{\nu_\tau}$  & $N_{\nu_\mu}$ & $N_{\nu_{\rm all}} $ & $N_\tau/N_{\nu_{\rm all}}$ \\
\cline{2-3} \cline{4-9}
$(1)$ &~90-95 & 0.14~&~60-90 & 0.059 & 0.049 &  0.011  & 0.12~& 1.14 \\
$(2)$ &~90-95 & 0.15~&~60-90 & 0.059 & 0.049 & 0.096 & 0.11~& 1.33 \\
$(3)$ &~90-95 & 0.23 &~60-90 & 0.079 & 0.062 & 0.0123 & 0.15~& 1.53 \\
$(4)$ &~90-95 & 0.12 &~60-90 & 0.046 & 0.037 & 0.0080 & 0.091~& 1.33\\
$(1)$ &~90-95 & 0.14 &~75-90 & 0.027 & 0.031 & 0.0056 & 0.064~& 2.14 \\
$(2)$ &~90-95 & 0.15 &~75-90 & 0.026 & 0.029 & 0.0048 & 0.060~& 2.47 \\
$(3)$ &~90-95 & 0.23 &~75-90 & 0.036 & 0.041 & 0.0062 & 0.083~& 2.75\\
$(4)$ &~90-95 & 0.12 &~75-90 & 0.021 & 0.024 & 0.0040& 0.049~& 2.45\\
\hline
\end{tabular}
\end{table}

Hereafter we consider $\Phi_0^{\nu_\alpha} (E_\nu) \propto E_\nu^{-2}$
as our nominal spectrum. We then estimate systematic uncertainties
associated with: different assumptions of the spectrum shape,
different parton distribution functions (GRV92NLO~\cite{Gluck:1998js}
and CTEQ66c~\cite{Nadolsky:2008zw}), and different estimates on
$\beta_\tau$~\cite{Armesto:2007tg}.  The contribution of different
systematic errors are listed in Table~\ref{table3}.

\begin{table}[htb]
\vspace*{-0.1in}
\caption{Contributions to the systematic uncertainty on the Earth-skimming to quasi-horizontal event ratio. We have considered the energy range $9.5 < \log_{10} (E_\nu/{\rm GeV}) < 10.5$ and the zenith angle range $75^\circ < \theta < 90^\circ$.}
\label{table3}
\begin{tabular}{c|ccc|c}
\hline
~~~~ratio~~~~&~~~~flux~~~~&~~~~PDF~~~~&~~~~$\beta_\tau$~~~~~~&~~~~sum~~~~\\
\hline
  & $+ 11\% $ & $0\%$ & $+24\%$ & $\phantom{2.4666666} + 26\%$ \\
2.47 &  &  &  & 2.47 \\
   &$-13\%$ & $-21\%$ & $-25\%$ & $\phantom{2.4666666} - 35\%$\\
\hline
\end{tabular}
\end{table}

\section{Auger Discovery Reach} 
\label{sec:results}

Consider a flux of neutrinos with energy in the range $10^{9.5}~{\rm
  GeV}<E_\nu< 10^{10.5}~{\rm GeV}$.  In the SM, the interaction path
length is
\begin{equation}
L_{\rm CC}^{\nu} = \left[ N_A \rho_s \sigma^{\nu}_{\rm CC}
\right] ^{-1} \ ,
\end{equation}
where $\sigma^{\nu}_{\rm CC}$ is the charged current cross-section for
$E_{\nu} = E_0$.  (We neglect neutral current interactions, which at
these energies serve only to reduce the neutrino energy by
approximately 20\%, which is within the systematic uncertainty.)  For
$E_0 \sim 10^{10}~{\rm GeV}$, $L_{\rm CC}^{\nu} \sim {\cal
  O}(100)~\km$.  Supplemented by the possibility of new
non-perturbative physics, the interaction path length is
\begin{equation}
L_{\rm tot}^{\nu} = \left[ N_A \rho_s (\sigma^{\nu}_{\rm CC} +
\sigma^{\nu}_{\rm NP}) \right] ^{-1} \ ,
\end{equation}
where $\sigma^{\nu}_{\rm NP}$ is the new physics contribution to the
cross-section for $E_{\nu} = E_0$.

\begin{figure*}[tbp]
\begin{minipage}[t]{0.32\textwidth}
\postscript{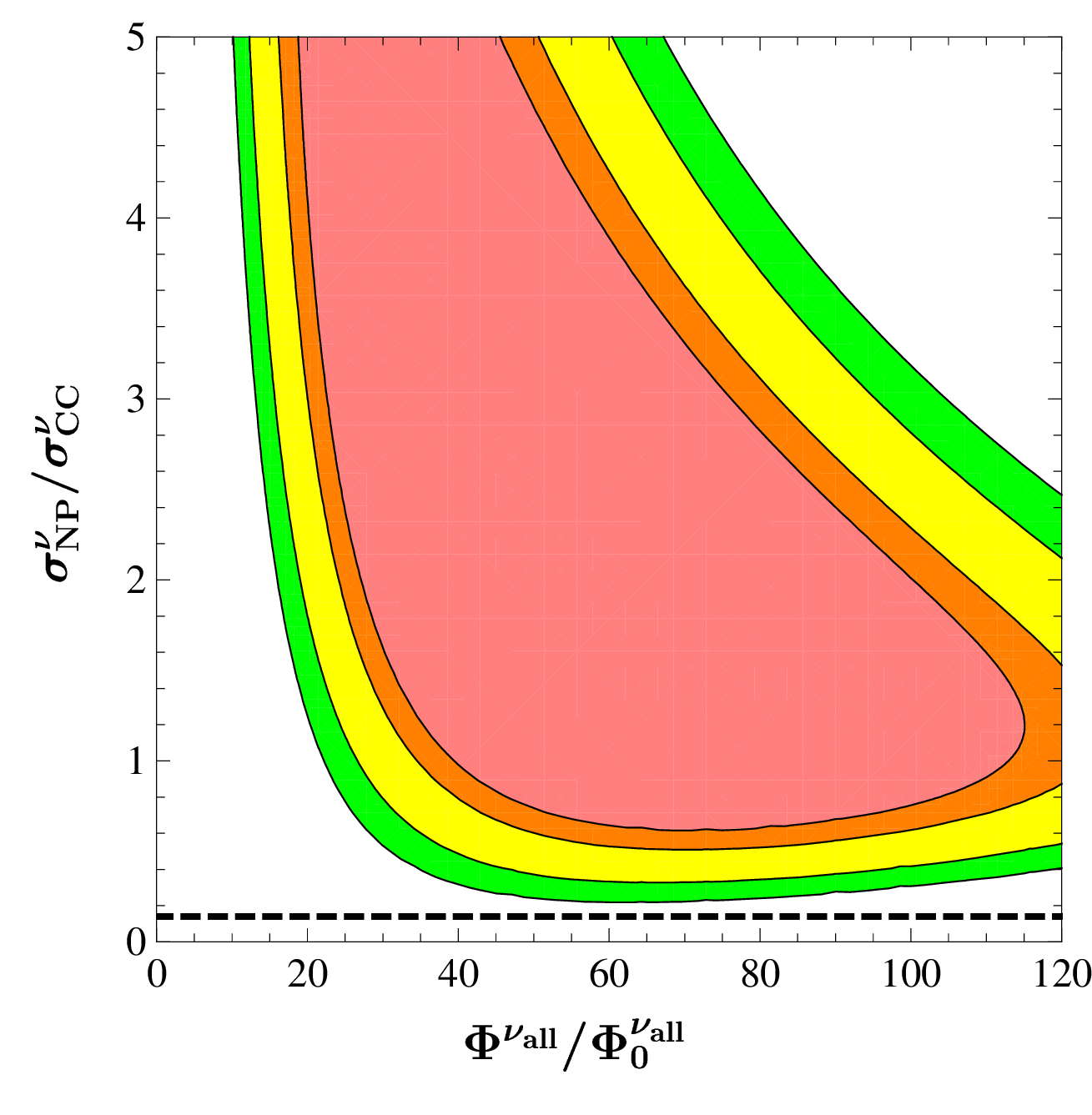}{0.99}
\end{minipage}
\hfill
\begin{minipage}[t]{0.32\textwidth}
\postscript{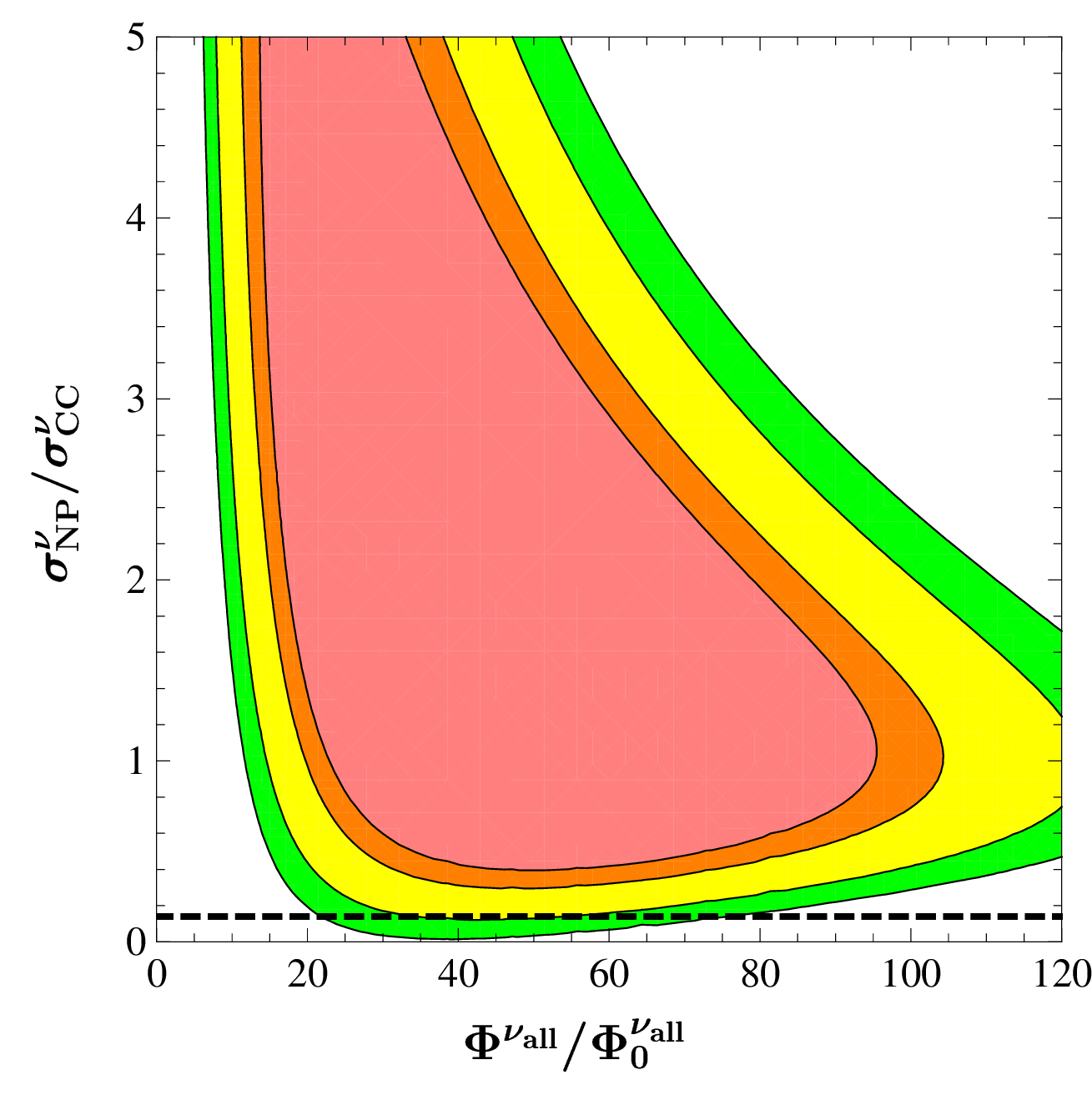}{0.99}
\end{minipage}
\hfill
\begin{minipage}[t]{0.32\textwidth}
\postscript{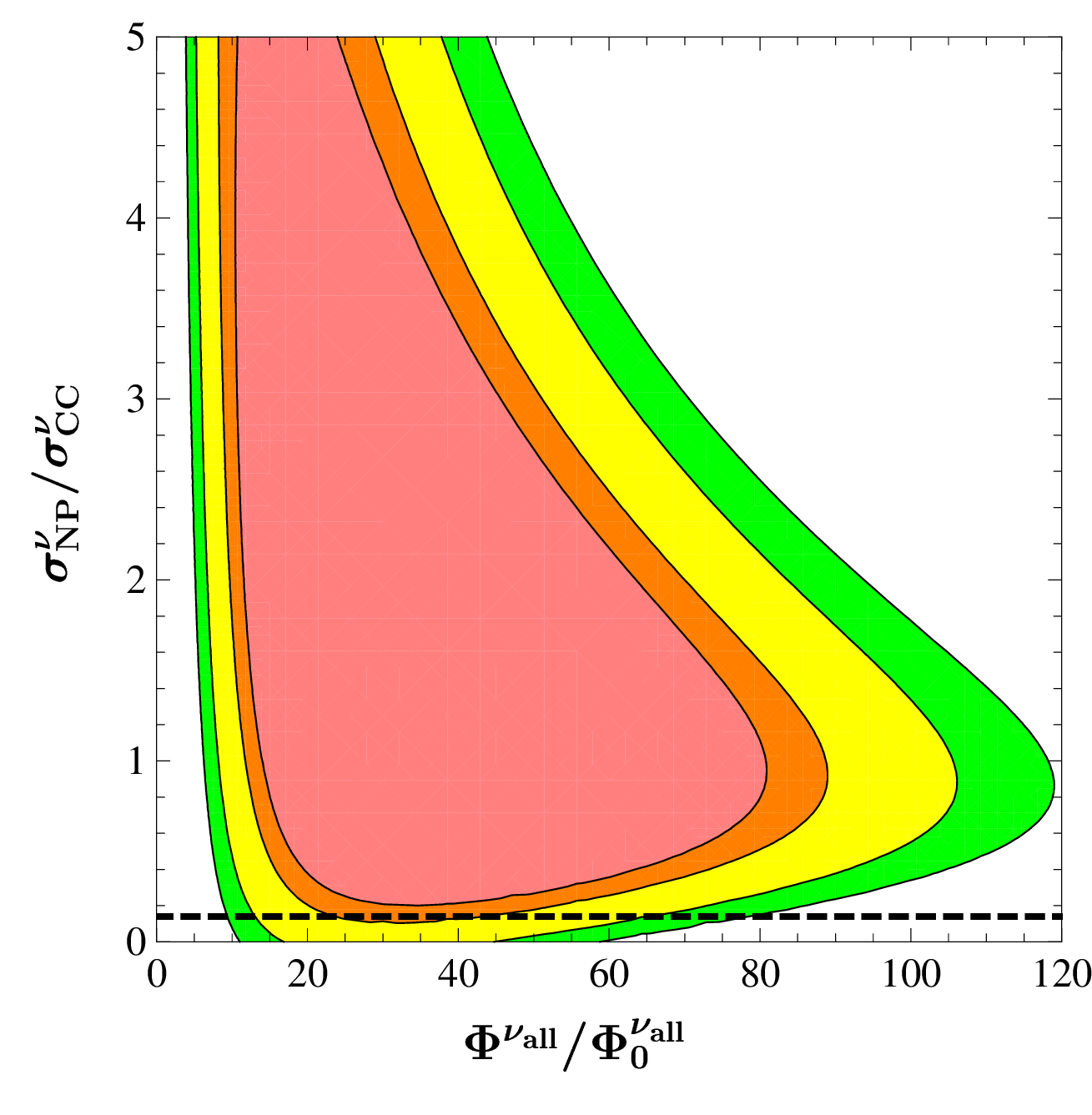}{0.99}
\end{minipage}
\caption{Projected determination of neutrino fluxes and cross-sections
  at $\sqrt{s} \approx 250$~TeV from future Auger data. The different
  shaded regions indicate the 90\%, 95\%, 99\% and 3$\sigma$
  confidence level contours in the $\Phi^{\nu_{\rm
      all}}/\Phi_0^{\nu_{\rm all}}-\sigma_{\rm NP}/\sigma_{\rm CC}$
  plane, for $N^{\rm obs}_{\rm ES} = 1$, $N^{\rm obs}_{\rm QH} = 10$
  (left), $N^{\rm obs}_{\rm ES} = 1$, $N^{\rm obs}_{\rm QH} = 7$
  (middle), and $N^{\rm obs}_{\rm ES} = 1$, $N^{\rm obs}_{\rm QH} = 5$
  (right). The dashed line indicates the result of including the
  systematic uncertainty on the NLO QCD CC neutrino-nucleon
  cross-section~\cite{Mandy}.}
\label{statistics}
\end{figure*}

The maximal path length for a detectable $\tau$ is given by
\begin{equation}
L^{\tau} = \frac{1}{\beta_{\tau} \rho_s} \ln \left( E_{\rm max} /
E_{\rm min} \right) \ ,
\label{ltau}
\end{equation}
where $E_{\rm max} \approx E_0$ is the energy at which the tau is
created, and $E_{\rm min}$ is the minimal energy at which a $\tau$ can
be detected.  For $E_{\rm max} / E_{\rm min} = 10$, $L^{\tau} =
11~\km$.

Given an isotropic $\nu_{\tau} + \bar{\nu}_{\tau}$ flux, the number of
taus that emerge from the Earth with sufficient energy to be detected
is proportional to an ``effective solid angle''
\begin{equation}
\Omega_{\rm eff} \equiv \int d\cos\theta\, d\phi\, \cos\theta\,
P(\theta,\phi) \ ,
\end{equation}
where
\begin{equation}
P(\theta,\phi) = \int_0^{\ell} \frac{dz}{L_{\rm CC}^{\nu}}
e^{-z/L_{\rm tot}^{\nu}} \
\Theta \left[ z - (\ell - L^{\tau} ) \right]
\label{P}
\end{equation}
is the probability for a neutrino with incident nadir angle $\theta$
and azimuthal angle $\phi$ to emerge as a detectable $\tau$. (In
Eq.\eqref{P}, for the reasons noted above, we have neglected the
possibility of detectable signals from new non-perturbative physics by
Earth-skimming neutrinos.)  Here $\ell = 2 R_{\oplus} \cos\theta$ is
the chord length of the intersection of the neutrino's trajectory with
the Earth, with $R_{\oplus} \approx 6371~\km$ the Earth's radius.
Evaluating the integrals, we find~\cite{Kusenko:2001gj}
\begin{eqnarray}
\Omega_{\rm eff} & = & 2 \pi
\frac{L_{\rm tot}^{\nu}}{L_{\rm CC}^{\nu}}
\left[ e^{L^{\tau} / L_{\rm tot}^{\nu}} - 1 \right] 
\left[ \left( \frac{L_{\rm tot}^{\nu}}{2 R_{\oplus}} \right)^2 \right. \nonumber \\
&- & \left. \left( \frac{L_{\rm tot}^{\nu}}{2 R_{\oplus}} +
\left( \frac{L_{\rm tot}^{\nu}}{2 R_{\oplus}} \right)^2 \right)
e^{-2R_{\oplus} / L_{\rm tot}^{\nu}} \right] \ .
\label{Omegaeff}
\end{eqnarray}
At the relevant energies, the neutrino interaction length satisfies
$L_{\rm tot}^{\nu} \ll R_{\oplus}$.  In addition, for $L_{\rm
  tot}^{\nu} \gg L^{\tau}$, valid when the cross-section enhancement
is significant but not so large as typical hadronic cross-section,
Eq.\eqref{Omegaeff} simplifies to~\cite{Anchordoqui:2001cg}
\begin{equation}
\Omega_{\rm eff} \approx
2\pi \frac{L_{\rm tot}^{\nu\, 2} L^{\tau}}{4 R_{\oplus}^2
L_{\rm CC}^{\nu}} \ .
\label{omegaeffapprox}
\end{equation}

Equation~(\ref{omegaeffapprox}) gives the functional dependence of the
Earth-skimming event rate on the new physics cross-section.  This rate is, of
course, also proportional to the source neutrino flux $\Phi^{\nu_{\rm all}}$ at
$E_0$. Given
these inputs,
\begin{equation}
N_{\rm ES} \approx
C_{\rm ES}\, \frac{\Phi^{\nu_{\rm all}}}{\Phi^{\nu_{\rm all}}_0}
\frac{\sigma^{\nu\, 2}_{\rm CC}}
{\left( \sigma^{\nu}_{\rm CC} + \sigma^{\nu}_{\rm NP} \right)^2} \ ,
\label{ES}
\end{equation}
where $C_{\rm ES} = 0.15$ is the number of Earth-skimming events
expected for a fiducial flux $\Phi^{\nu_{\rm all}}_0$ in the abscence
of new physics.

In contrast to Eq.\eqref{ES}, the rate for quasi-horizontal showers
has the form
\begin{equation}
N_{\rm QH} = C_{\rm QH} \frac{\Phi^{\nu_{\rm all}}}{\Phi^{\nu_{\rm all}}_0}
\frac{\sigma^{\nu}_{\rm CC} + \sigma^{\nu}_{\rm NP}}
{\sigma^{\nu}_{\rm CC}} \ ,
\label{QH}
\end{equation}
where $C_{\rm QH} = 0.06$ for the Auger Surface Array, as determined
in Sec.~\ref{sec:sims}.

Given a flux $\Phi^{\nu_{\rm all}}$ and new physics cross-section
$\sigma^{\nu}_{\rm NP}$, both $N_{\rm ES}$ and $N_{\rm QH}$ are
determined.  On the other hand, given just a quasi-horizontal event
rate $N_{\rm QH}$, it is impossible to differentiate between an
enhancement of the cross-section due to new physics and an increase on
the flux.  However, in the region where significant event rates are
expected the contours of $N_{\rm QH}$ and $N_{\rm ES}$, given by
Eqs.~(\ref{ES}) and (\ref{QH}), are more or less orthogonal and
provide complementary information.  With measurements of $N_{\rm
  QH}^{\rm obs}$ and $N_{\rm ES}^{\rm obs}$, both $\sigma^{\nu}_{\rm
  NP}$ and $\Phi^{\nu_{\rm all}}$ may be determined independently, and
neutrino interactions beyond the SM may be unambiguously identified.

We now turn to determining the projected sensitivity of Auger to
neutrino fluxes and cross-sections. The quantities $N_{\rm ES}$ and
$N_{\rm QH}$ as defined in Eqs.~(\ref{ES}) and (\ref{QH}) can be
regarded as the theoretical values of these events, corresponding to
different points in the $\Phi^{\nu_{\rm all}}/\Phi_0^{\nu_{\rm
    all}}-\sigma_{\rm NP}/\sigma_{\rm CC}$ parameter space. For a
given set of observed rates $N^{\rm obs}_{\rm ES}$ and $N^{\rm
  obs}_{\rm QH}$, two curves are obtained in the two-dimensional
parameter space by setting $N^{\rm obs}_{\rm ES} = N_{\rm ES}$ and
$N^{\rm obs}_{\rm QH} = N_{\rm QH}$.  These curves intersect at a
point, yielding the most probable values of flux and cross section for
the given observations.  Fluctuations about this point define contours
of constant $\chi^2$ in an approximation to a multi-Poisson likelihood
analysis. The contours are defined by
\begin{equation}
\chi^2  =  \sum_i 2 \,\left[ N_i  -  N_i^{\rm obs}\right] +
2\,  N_i^{\rm obs}\,
\ln \left[ N_i^{\rm obs}/ N_i \right] \,\,,
\label{baker}
\end{equation} 
where $i=$~ES, QH~\cite{Baker:1983tu}. In Fig.~\ref{statistics}, we
show results for three representative cases.  Assuming ($N^{\rm
  obs}_{\rm ES} = 1$, $N^{\rm obs}_{\rm QH} = 10$), ($N^{\rm obs}_{\rm
  ES} = 1$, $N^{\rm obs}_{\rm QH} = 7$), and ($N^{\rm obs}_{\rm ES} =
1$, $N^{\rm obs}_{\rm QH} = 5$) we show the 90\%, 95\%, 99\% and
3$\sigma$ CL contours for 2 d.o.f. ($\chi^2 = 4.61,\ 5.99,\ 9.21,$ and
11.83, respectively).  For $N^{\rm obs}_{\rm ES} = 1$ and $N^{\rm
  obs}_{\rm QH} = 10$, the possibility of a SM interpretation along
the $\sigma^{\nu}_{\rm NP} = 0$ axis (taking into account systematic
uncertainties) would be excluded at greater than 99\% CL for {\em any}
assumed flux. The power of the Earth-skimming information is such that
the best fit consistent with the SM would require a flux of about 50
times the Waxman-Bahcall flux, which is already excluded by present
limits~\cite{Anchordoqui:2009nf}.

\section{Summary}\label{summary}
                                                                               
We have re-examined a technique to search for new physics at sub-fermi
distances. The strategy involves determining the ratio of
quasi-horizontal to Earth-skimming showers initiated by cosmic
neutrinos which would need to be detected by the Pierre Auger
Observatory in order to signal the existence of exotic
non-perturbative interactions beyond the TeV-scale. We perform Monte
Carlo simulations of neutrino interactions in the Earth and in the
atmosphere, and realistic simulation of the detector acceptance using
the Auger \Offline software.  We find that observation of 1
Earth-skimming and 10 quasi-horizontal events would exclude the
standard model at the 99\% confidence level.  If new non-perturbative
physics exists, a decade or so would be required to uncover it in the
most optimistic case (cosmic neutrino flux at the Waxman-Bahcall level
and $\nu N$ cross-section about an order of magnitude above the
standard model prediction). The proposed Northern Auger
site~\cite{Abraham:2009en} (which has not been optimized for neutrino
studies) would reduce this time by about a factor of 2. Any hint of
such an important signal would provide an impetus to infill the array
to increase the neutrino acceptance.
  
\section*{Acknowledgements}

We would like to thank Mandy Cooper-Sarkar for discussions. L.A.A.\ is
supported by the U.S. National Science Foundation (NSF) Grant No
PHY-0757598, and the UWM RGI.  H.G.\ is supported by the NSF Grant No
PHY-0757959. D.G.\ is supported by the HHNG-128 grant of the Helmholtz
association and the Polish Ministry of Science and Higher Education
under Grant 2008 No. NN202 127235. T.P.\ is supported by the NSF Grant
No PHY-0855388. M.R.\ is supported by the HHNG-128 grant of the
Helmholtz association. S.S.\ acknowledges support by the EU Marie
Curie Network ``UniverseNet'' (HPRN-CT-2006-035863).  L.L.W.\ is
supported by the UWM RGI. Any opinions, findings, and conclusions or
recommendations expressed in this material are those of the authors
and do not necessarily reflect the views of the National Science
Foundation.

\end{document}